# Turbulent kinetic energy in 2D isothermal interchange-dominated scrape-off layer ExB drift turbulence: governing equation and relation to particle transport

R. Coosemans,[1, a)] W. Dekeyser,[1, b)] and M. Baelmans[1, c)]
*Applied Mechanics and Energy Conversion Section, Department of Mechanical Engineering, KU Leuven, Leuven 3001, Belgium*

(Dated: 30 November 2020)

This paper studies the turbulent kinetic energy ($k_\perp$) in 2D isothermal electrostatic interchange-dominated ExB drift turbulence in the scrape-off layer, and its relation to particle transport. An evolution equation for the former is analytically derived from the underlying turbulence equations. Evaluating this equation shows that the dominant source for the turbulent kinetic energy is due to interchange drive, while the parallel current loss to the sheath constitutes the main sink. Perpendicular transport of the turbulent kinetic energy seems to play a minor role in the balance equation. Reynolds stress energy transfer also seems to be negligible, presumably because no significant shear flow develops under the given assumptions of isothermal sheath-limited conditions in the open field line region. The interchange source of the turbulence is analytically related to the average turbulent ExB energy flux, while a regression analysis of TOKAM2D data suggests a model that is linear in the turbulent kinetic energy for the sheath loss. A similar regression analysis yields a diffusive model for the average radial particle flux, in which the anomalous diffusion coefficient scales with the square root of the turbulent kinetic energy. Combining these three components, a closed set of equations for the mean-field particle transport is obtained, in which the source of the turbulence depends on mean flow gradients and $k_\perp$ through the particle flux, while the turbulence is saturated by parallel losses to the sheath. Implementation of this new model in a 1D mean-field code shows good agreement with the original TOKAM2D data over a range of model parameters.

## I. INTRODUCTION

The anomalous transport observed in tokamaks is generally known to be caused by turbulent fluctuations. ExB drift turbulence, fluctuating drift flows caused by electric field fluctuations, is believed to be dominant in the plasma edge[1–3]. Usually, the interchange instability provides the main source of this turbulence in the scrape off layer (SOL), with the drift wave mechanism playing a less important role as perturbations without a parallel component can exist in this open field-line region[4–6].

Various approaches have been followed to simulate this ExB drift turbulence and the related particle and heat transport, ranging from gyrokinetic models (e.g. GENE[7]) to fluid turbulence codes (e.g. TOKAM2D[8–10] and TOKAM3X[11]). However, the high computational cost of these approaches inhibits their use for simulations of complete future fusion reactors with the existing computational resources. Hence, mean-field plasma edge transport codes (e.g. SOLPS-ITER[12]), which effectively calculate averaged quantities of the flow only, are expected to remain the main instruments for the design of divertors for future reactors. However, these transport codes do not resolve the turbulence, but instead use ad hoc transport coefficients to model the resulting turbulent transport[13–15], limiting the predictive capacities of these codes. One approach to remedy this deficiency in mean-field transport codes is to couple them to turbulence codes that can then provide transport coefficients self-consistently. In such an approach, the turbulence code can either resolve the turbulence locally in distinct parts of the mean-field domain as performed by for example Nishimura *et al.*[16], or it can globally solve the entire mean-field computational domain as performed by Zhang *et al.*[17]. Note that in the former approach non-local transport effects cannot be captured, while in the latter approach the computational cost is expected to remain a bottleneck due to the need for a 3D turbulence simulation resolving the relevant turbulent length and time scales.

Also in hydrodynamic turbulence modelling, there is a wide gap between detailed turbulent simulations using Direct Numerical Simulation (DNS), and the application to flows in realistic configurations. A range of techniques has been developed to bridge this gap. In DNS, the Navier-Stokes equations are solved on very short length and time scales, such that all relevant scales of the turbulence are resolved. In Large Eddy Simulations (LES), only the larger scales of the flow are resolved, while the smaller scales are filtered out and modelled using a subgrid model instead. In the Reynolds-Averaged Navier-Stokes (RANS) approach, a decomposition in time-averaged and fluctuating components leads to transport equations for the mean-field quantities, complemented with models for nonlinear closure terms[18].

Recently, Bufferand *et al.*[19] proposed a mean-field model for the turbulent particle transport in the plasma edge that draws inspiration from these RANS models. More specifically, the model bears similarity to $k(-\varepsilon)$ models, where equations for the turbulent kinetic energy $k$ (and dissipation $\varepsilon$) are solved to provide time- and length scales to model the closure terms[18]. Bufferand *et al.* proposed a diffusive model for the radial particle transport where the anomalous trans-

---

a)Electronic mail: reinart.coosemans@kuleuven.be
b)Electronic mail: wouter.dekeyser@kuleuven.be
c)Electronic mail: martine.baelmans@kuleuven.be

port coefficients scale linearly with the local turbulent kinetic energy. Bufferand's equation for the turbulent kinetic energy is based on a predator-prey model for the turbulence intensity derived by Miki *et al.*[20] with some ad hoc adaptations to obtain a transport model for the turbulent kinetic energy. This model has later been refined by using global confinement scaling laws in its closure of the dissipation[21–23].

Similar to Bufferand *et al.*, the present paper relates the average turbulent particle transport to turbulent quantities and the turbulent kinetic energy in particular. In addition, in this study we elaborate the equations governing the transport and the evolution of $k_\perp$ in a consistent analytical way. Fluid turbulence codes, which resolve all the fine length- and time scales of the flow (as DNS codes do for hydrodynamic turbulence), are then used to explore the physics that need to be included in the mean-field models and to propose models for the remaining closure terms. More specifically, the isothermal TOKAM2D turbulence code[9,10] for the scrape-off layer (SOL) is used as reference in our study. As such, 2D, isothermal, quasi-neutral, electrostatic plasmas will be investigated. The turbulence is assumed to be dominated by the interchange instability and mean quantities will only vary in the radial direction, resulting in 1D averaged profiles and transport.

The remainder of this paper is structured as follows. Section II introduces the model describing 2D isothermal interchange turbulence, as implemented in the TOKAM2D code. This code is used as a DNS reference in this work. Next, analytical equations for the total, turbulent and mean-flow kinetic energy are derived in section III. Section IV evaluates the turbulent kinetic energy equation using TOKAM2D data and investigates the balance of the sources and sinks of turbulent kinetic energy. Models for the dominant closure terms are proposed based on a regression analysis. The same regression analysis techniques are used to identify possible models for the average turbulent particle flux in section V. Section VI then compares simulation results obtained with the newly developed model to the original TOKAM2D results and to the model proposed by Bufferand *et al.* Finally, section VII summarises the main findings of this work and comments on suggestions for further research.

## II. MODEL FOR 2D ISOTHERMAL INTERCHANGE TURBULENCE

The 2D interchange turbulence model in TOKAM2D[8–10] consists of the continuity equation, the vorticity equation and the corresponding definition of the vorticity:

$$\frac{\partial n}{\partial t} + [\phi,n] = S_n - \sigma_N c_s n \exp(\Lambda - \frac{\phi}{T_e}) + D_n \nabla_\perp^2 n, \quad (1)$$

$$\frac{\partial \omega}{\partial t} + [\phi,\omega] = \frac{1}{n}[nT, gx]$$
$$+ \sigma_W c_s (1 - \exp(\Lambda - \frac{\phi}{T_e})) + \nu \nabla_\perp^2 \omega, \quad (2)$$

$$\omega = \nabla_\perp^2 \phi. \quad (3)$$

These equations are written for a Cartesian coordinate system where $\parallel$ denotes the parallel direction along the magnetic unit vector **b**, $x$ is the radial direction, and $y$ is perpendicular to the other two, such that the three form a right handed coordinate system. Hence, $y$ points in the electron diamagnetic direction. In equations 1-3, $n$ is the density, $\phi$ the electrostatic potential, $S_n$ the particle source, $\sigma_N$ and $\sigma_W$ parameters quantifying the magnitude of the sheath losses ($\sigma \sim 1/L_\parallel$ with $L_\parallel$ the parallel connection length), $c_s$ the sound speed, $\Lambda$ the sheath potential, $T_e$, $T_i$ and $T = T_e + T_i$ the (constant) electron, ion and total temperatures respectively, $D_n$ a diffusion constant, $\omega$ the vorticity, $g$ a parameter characterising the radial decay of the magnetic field ($g \sim 1/R$ with $R$ the major radius of the tokamak) and $\nu$ a viscosity. The Poisson bracket in these equations is defined as $[P,Q] = \mathbf{b} \cdot (\nabla P \times \nabla Q)$. Note that all quantities in these equations are normalised to the reference gyro-frequency $\Omega_{ref} = qB_{ref}/m$ and gyro-radius $\rho_{ref} = \Omega_{ref}^{-1}\sqrt{T_{ref}/m}$, where $q$ is the ion charge, $m$ the ion mass and $B_{ref}$ the magnetic field strength. All equations in the remainder of this paper will follow this normalisation, unless specifically mentioned otherwise.

In deriving this set of equations it is assumed that the ExB drift dominates the perpendicular velocity such that it is the only component that needs to be taken into account in the convective terms and in the polarisation current term. The convective operator for an arbitrary quantity $u$ takes the form $[\phi,u] = \mathbf{V}_E \cdot \nabla u$, where $\mathbf{V}_E$ is the ExB drift velocity. However, the definition of the Poisson bracket implies that $V_E$ is calculated as $\mathbf{V}_E = \mathbf{b} \times \nabla \phi$, hence without taking variations of the magnetic field strength into account as is normally done, i.e. $\mathbf{V}_E = (\mathbf{b} \times \nabla \phi)/B$. As a result, $\nabla \cdot V_E = 0$, such that the convective operator $[\phi,u]$ may equivalently be written as $\nabla \cdot (u\mathbf{V}_E)$. We will write the equations in this paper in the conservative form using the divergence operator to allow easier generalisation to more complex models. In equation 2 the Boussinesq approximation is also made, assuming that $\nabla \cdot \mathbf{J}_p = \nabla \cdot (qn\mathbf{V}_p) \approx qn\nabla \cdot (\mathbf{V}_p)$, where $\mathbf{V}_p = b \times (\frac{\partial \mathbf{V}_E}{\partial t} + \mathbf{V}_E \cdot \nabla \mathbf{V}_E)$ and $\mathbf{J}_p = n\mathbf{V}_p$ are the polarisation velocity and current density respectively. The parallel direction is modelled by using analytical relations for the sheath behaviour, assuming every 2D cell to be connected to the sheath in the unresolved third direction. The particle source $S_n$ has a Gaussian profile in the (radial) $x$-direction and is constant both in the (diamagnetic) $y$-direction and in time. In the simulations performed here, the particle source is situated towards the inner boundary of the domain.

The equations 1-3 are solved in 2D (radial-diamagnetic) by the finite volume version of the TOKAM2D code[9,10]. Periodic boundary conditions are used on diamagnetic ($y$) boundaries, while Neumann boundary conditions are applied on radial ($x$) boundaries such that there is no radial flux of any quantity at radial boundaries of the domain. In addition, fringe regions are applied near the radial boundaries of the domain to drive fluctuations in the $y$-direction to zero in those regions. Note that the data used further on in this manuscript only considers the physical middle part of the computational domain where there is no fringe region and the influence of the particle source is negligible.

As mentioned before, we will try to find a mean-field model for the turbulent transport, drawing inspiration from RANS

techniques for hydrodynamic turbulence. To this end, all turbulent quantities in the governing equations 1-3 are split into a mean flow and a fluctuating component. Two types of decompositions are used for this: Reynolds decomposition and Favre decomposition. The Reynolds decomposition is defined as follows[18,24]:

$$u = \bar{u} + u', \quad (4)$$

where the mean-field quantity $\bar{u}$ is the ensemble average of the quantity $u$ and $u'$ its fluctuating component. The ensemble average is the average over an infinite amount of realisations of the the turbulent flow. Then, we assume the turbulent flow to be ergodic such that a long time statistical steady state of the flow exists, of which the time average converges to this ensemble average:

$$\bar{u} = \lim_{T \to \infty} \int_0^T u \, dt, \quad (5)$$

with $t$ being time.

The Favre or density-weighted average[24], is defined as

$$u = \tilde{u} + u'',$$
$$\tilde{u} = \frac{\overline{nu}}{\bar{n}}. \quad (6)$$

This density-weighted average is introduced because it appears naturally in variable density flows[24]. These definitions imply the following relationships, which will be frequently used in the derivations below:

$$\overline{u'} = 0$$
$$\overline{nu''} = 0. \quad (7)$$

Note also that the time-averaging operator $\bar{u}$ commutes with time and space derivatives, but the Favre operator $\tilde{u}$ does not.

Given these decompositions, we will start our analysis by time-averaging the continuity and vorticity equations 1-3 to obtain equations describing the evolution of the time-averaged density $\bar{n}$, vorticity $\bar{\omega}$ and electrostatic potential $\bar{\phi}$:

$$\frac{\partial \bar{n}}{\partial t} + [\bar{\phi}, \bar{n}] + \overline{[\phi', n']}$$
$$= \bar{S}_n - \sigma_n c_s n \exp(\Lambda - \frac{\phi}{T_e}) + D_n \nabla_\perp^2 \bar{n}, \quad (8)$$

$$\frac{\partial \bar{\omega}}{\partial t} + [\bar{\phi}, \bar{\omega}] + \overline{[\phi', \omega']} = \overline{\frac{1}{n}[nT, gx]}$$
$$+ \sigma_W c_s(1 - \overline{\exp(\Lambda - \frac{\phi}{T_e})}) + \nu \nabla_\perp^2 \bar{\omega}, \quad (9)$$

$$\bar{\omega} = \nabla_\perp^2 \bar{\phi}. \quad (10)$$

The terms in these equation can be compared to the corresponding terms in the turbulent equations 1-3. Linear terms, such as the time derivative and diffusive terms, retain the same form as in the original equation. Nonlinear terms, such as the convective terms, the sheath loss terms and the interchange term, lead to correlations between fluctuations, and give rise to terms that require closure. Note that we opt to use a Reynolds average for the potential since this clearly presents the expected relation between the mean-field ExB drift and the mean-field potential. Finally, $\bar{S}_n$ represents the time-averaged source term of particles. As the particle source in the TOKAM2D code is constant in time (and uniform in the y direction), this overbar notation is not strictly necessary, but is used for generality and possible future extensions of the model.

Since the diamagnetic *y*-direction in TOKAM2D is periodic, averaged quantities only vary in the radial *x*-direction. For this reason, the TOKAM2D data used in the next sessions will not only be averaged over time as indicated in the average in equation 5, but also over this periodic *y*-direction. Note that this diamagnetic averaging is not a necessary part of the methodology, but is only used to obtain more data points for the averaging. The mean field models for the turbulence that will be developed in this work will be 1D, radial models. As a result, only the turbulent ExB particle flux $\bar{\Gamma}_{E,t} = \overline{n' \mathbf{V}_E'}$ contributes to the averaged (radial) total ExB particle flux because the mean field ExB particle flux $\bar{\Gamma}_{E,m} = \bar{n} \bar{\mathbf{V}}_E$ is zero (since gradients in the *y*-direction of averaged quantities are zero). For this reason the term $[\bar{\phi}, \bar{n}] = \nabla \cdot \bar{\Gamma}_{E,m}$ drops from equation 8. Hence, the remainder of this paper will look for models to close the turbulent ExB particle flux $\bar{\Gamma}_{E,t} = \overline{n' \mathbf{V}_E'}$ in mean-field transport models. Note however that in more general (non-1D) models, solving equations 9 and 10 for $\bar{\omega}$ and $\bar{\phi}$ would allow to calculate the mean-field ExB particle transport $\bar{\Gamma}_{E,m}$. Note also that this would require additional closures for the nonlinear terms in the averaged vorticity equation as well.

### III. DERIVATION OF $k_\perp$ EQUATION

As section II has shown the average particle transport in the 2D interchange turbulence model to be governed by the correlations between density and potential fluctuations, we aim to find a measure for the intensity of these fluctuations, and relate it to the resulting particle transport. To this end, we define the total ($E_{k,\perp}$), mean flow ($E_{k,mean,\perp}$), and turbulent ($k_\perp$) perpendicular kinetic energies as

$$E_{k,\perp} = \frac{\mathbf{V}_E^2}{2}, \quad (11)$$

$$\bar{n} E_{k,mean,\perp} = \frac{\bar{n} \tilde{\mathbf{V}}_E^2}{2}, \quad (12)$$

$$\bar{n} k_\perp = \frac{\overline{n \mathbf{V}_E''^2}}{2}. \quad (13)$$

The symbol $k_\perp$ here is not to be confused with the perpendicular wave number, which in plasma physics literature is often denoted with this symbol as well. Note that $E_{k,\perp}$ varies rapidly in time and space as it follows the instantaneous fluctuations, while $E_{k,mean,\perp}$ and $k_\perp$ are time averaged quantities that do not change at these small scales. The latter two are constant in time in a statistical steady state, while the former is not. Note also that the sum of mean flow and turbulent kinetic energy per unit volume equals the averaged total kinetic energy per unit volume:

$$\overline{n E_{k,\perp}} = \bar{n} E_{k,mean,\perp} + \bar{n} k_\perp. \quad (14)$$

The turbulent kinetic energy as defined in equation 13 provides a direct measure of the characteristic (density weighed) ExB drift velocity of ions in the fluctuating electrostatic field. This is exactly the motion that is believed to cause the anomalous transport observed in the SOL that is of interest in this paper[1–3,25]. This close link between the average radial particle flux and the perpendicular turbulent kinetic energy $k_\perp$ will be confirmed by TOKAM2D data in section V.

In the present section, we analytically derive the equation governing the transport of $k_\perp$ from the vorticity equation. This will provide some insight in the physics of the turbulence and the mechanisms for its transport, creation and destruction. First, an equation for the total kinetic energy is derived in subsection III A. Then equations for the mean flow kinetic energy and the turbulent kinetic energy are derived in III B. We follow a procedure similar to Scott[1], Garcia *et al.*[26] and Tran *et al.*[27], but rigorously accounting for density fluctuations in the kinetic energy equation. Note that taking these density fluctuations into account in the $k_\perp$ equation may not play a very large role in the results of the present analysis of incompressible flow. However, we decided not to neglect density fluctuation a priori, for generality, and in anticipation of future extensions of the model presented in this manuscript where they may be more important. Retaining these density fluctuations does render the derivation and the final expressions more complicated though, but the Favre averages demonstrated in equation 6 allow to reduce the number of closure terms that appear in the mean-field equations with respect to using Reynolds averages shown in equation 4. Similar Favre averaging techniques have been used to take density fluctuations into account to analyse zonal flow generation by Held *et al.*[28].

### A. Total kinetic energy equation

To obtain an expression for the time rate of change of the total kinetic energy, we multiply the divergence of the polarisation velocity $V_p$ with $n\phi$ and use the continuity equation 1 to rewrite:

$$\frac{\partial}{\partial t}nE_{k,\perp} + \nabla \cdot (nE_{k,\perp}\mathbf{V}_E)$$
$$= n\phi \nabla \cdot \mathbf{V}_p - n\nabla \cdot (\phi \mathbf{V}_p) + E_{k,\perp}S_n. \quad (15)$$

In expression 15 we insert the vorticity equation 2, which is effectively a charge balance equation with the left hand side representing $-\nabla \cdot V_p$, to find the total kinetic energy equation:

$$\frac{\partial}{\partial t}nE_{k,\perp} + \nabla \cdot (\Gamma_{E_{k_\perp}})$$
$$= -\phi[nT, gx] - \sigma_W c_s n\phi (1 - \exp(\Lambda - \frac{\phi}{T_e}))$$
$$- \nu n\phi \nabla_\perp^2 \omega + \phi \mathbf{V}_p \cdot \nabla n + S_{E_{k_\perp},n}, \quad (16)$$

$$\Gamma_{E_{k_\perp}} = nE_{k,\perp}\mathbf{V}_E + \phi \mathbf{J}_p, \quad (17)$$

$$S_{E_{k_\perp},n} = E_{k,\perp}S_n + D_n E_{k,\perp}\nabla_\perp^2 n$$
$$- E_{k,\perp}\sigma_N c_s n \exp(\Lambda - \frac{\phi}{T_e}). \quad (18)$$

The terms on the LHS of this equation represent the time rate of change and transport of $k_\perp$, with the transport terms written in conservative form. In equation 16, only the ExB velocity appears in the divergence terms on the LHS. The contributions of the divergence of the parallel and diffusive flow components appear on the RHS of the equation, as part of the source term 18, consistent with the notation in equations 1 and 2. In more complete models, these two flow components would naturally be moved to the transport term on the LHS of the equation to ensure particle and energy conservation. The RHS of the equation groups sources and sinks of $k_\perp$. The first two are the interchange source and loss to the sheath through divergence of the parallel current, which will turn out to be the dominant ones. The following term is a dissipation term due to the viscosity. The second but last term on the RHS is a "Boussinesq correction term" introduced by bringing $n$ in the divergence in $n\nabla \cdot (\phi \mathbf{V}_p)$. This term would not have been present if the Boussinesq approximation had not been made in the vorticity equation 2.

### B. Mean flow and turbulent kinetic energy equations

In order to arrive at equations for $E_{k,mean,\perp}$ and $k_\perp$ defined in equations 12 and 13, the $E_{k,\perp}$ equation 16 should be split in a contribution due to mean flows and a contribution due to fluctuations.

We obtain an expression for the time rate of change of $E_{k,mean,\perp}$ by taking the scalar product of the average polarisation current $\bar{\mathbf{J}}_p$ and the Favre-averaged gradient of the electrostatic potential $\widetilde{\nabla \phi}$ and then using the averaged continuity equation 8 to rewrite:

$$\frac{\partial}{\partial t}\bar{n}E_{k,mean,\perp} + \nabla \cdot (\bar{n}\tilde{\mathbf{V}}_E E_{k,mean,\perp} + \overline{n\mathbf{V}''_E\mathbf{V}''_E} \cdot \tilde{\mathbf{V}}_E)$$
$$= -\widetilde{\nabla \phi} \cdot \bar{\mathbf{J}}_p + \overline{n\mathbf{V}''_E\mathbf{V}''_E} : (\nabla \tilde{\mathbf{V}}_E)^T$$
$$+ \tilde{\mathbf{V}}_E \cdot \overline{\mathbf{V}''_E S_n} + E_{k,mean,\perp}\bar{S}_n. \quad (19)$$

In the derivation of this equation, only the ExB velocity is considered to be important for the polarisation current.

In order to use this expression starting from an averaged charge balance equation ($\nabla \cdot \bar{J} = 0$), $\widetilde{\nabla \phi} \cdot \bar{J}_p$ is rewritten to include $\nabla \cdot \bar{\mathbf{J}}_p$:

$$-\widetilde{\nabla \phi} \cdot \bar{\mathbf{J}}_p = \bar{\phi}\nabla \cdot \bar{\mathbf{J}}_p - \nabla \cdot (\bar{\phi}\bar{\mathbf{J}}_p) - \frac{\bar{\mathbf{J}}_p}{\bar{n}} \cdot \overline{n'\nabla \phi'}. \quad (20)$$

Note that this is more complicated than for the total kinetic energy case because Favre averages and gradients do not commute. As a result, an additional "Favre averaging term", which is the last term in 20, originates. A relation between the time change of $E_{k,mean,\perp}$ and the averaged charge balance equation is finally found by inserting equation 20 in equation

Turbulent kinetic energy in 2D isothermal interchange-dominated scrape-off layer ExB drift turbulence 5

19:

$$\frac{\partial}{\partial t}\bar{n}E_{k,mean,\perp}$$
$$+\nabla\cdot(\bar{n}\tilde{\mathbf{V}}_E E_{k,mean,\perp}+\overline{n\mathbf{V}''_E\mathbf{V}''_E}\cdot\tilde{\mathbf{V}}_E+\bar{\phi}\bar{\mathbf{J}}_p)$$
$$=\bar{\phi}\nabla\cdot\bar{\mathbf{J}}_p+\overline{n\mathbf{V}''_E\mathbf{V}''_E}:(\nabla\tilde{\mathbf{V}}_E)^T$$
$$+\tilde{\mathbf{V}}_E\cdot\overline{\mathbf{V}''_E S_n}+E_{k,mean,\perp}\bar{S}_n-\frac{\bar{J}_p}{\bar{n}}\overline{n'\nabla\phi'} \quad (21)$$

Relation 21 can now be applied to the vorticity equation 2 to obtain the $E_{k,mean,\perp}$ equation:

$$\frac{\partial}{\partial t}\bar{n}E_{k,mean,\perp}+\nabla\cdot(\bar{\Gamma}_{E_{k,mean,\perp}})$$
$$=-\bar{\phi}\overline{[nT,gx]}-\sigma_W\bar{\phi}c_s n(1-\exp(\Lambda-\frac{\phi}{T_e}))$$
$$-\bar{\phi}\overline{n\nu\nabla^2_\perp\omega}+\overline{n\mathbf{V}''_E\mathbf{V}''_E}:(\nabla\tilde{\mathbf{V}}_E)^T$$
$$-\frac{\bar{\mathbf{J}}_p}{\bar{n}}\overline{n'\nabla\phi'}+\bar{\phi}\overline{\mathbf{V}_p\cdot\nabla n}+S_{E_{k,mean,\perp},n}, \quad (22)$$
$$\bar{\Gamma}_{E_{k,mean,\perp}}=\bar{n}\tilde{\mathbf{V}}_E E_{k,mean,\perp}+\overline{n\mathbf{V}''_E\mathbf{V}''_E}\cdot\tilde{\mathbf{V}}_E+\bar{\phi}\bar{\mathbf{J}}_p, \quad (23)$$
$$S_{E_{k,mean,\perp},n}=E_{k,mean,\perp}\bar{S}_n+\tilde{\mathbf{V}}_E\cdot\overline{\mathbf{V}''_E S_n}$$
$$+D_n E_{k,mean,\perp}\nabla^2_\perp\bar{n}+D_n\tilde{\mathbf{V}}_E\cdot\overline{\mathbf{V}''_E\nabla^2_\perp n}$$
$$-\sigma_n c_s E_{k,mean,\perp}n\exp(\Lambda-\phi/T_e)$$
$$-\sigma_N c_s\overline{n\mathbf{V}''_E\exp(\Lambda-\phi/T_e)}\cdot\tilde{\mathbf{V}}_E \quad (24)$$

In this derivation, vorticity equation 2 was multiplied by the density and averaged, and the relation $-\bar{\phi}\overline{n\nabla\cdot\mathbf{V}_p}=\bar{\phi}\nabla\cdot\bar{\mathbf{J}}_p-\bar{\phi}\overline{\mathbf{V}_p\cdot\nabla n}$ was used, leading to the appearance of a "Boussinesq correction term". Also, the parallel and diffusive particle fluxes are treated like volumetric particle sink as was done before, in section III A. Note that the last four terms in $S_{E_{k,mean,\perp},n}$ are still pure transport terms. Due to symmetry, the parallel contributions to the Reynolds stresses vanish.

We find the $k_\perp$ equation by taking the difference of the average of the $E_{k,\perp}$ equation (average of equation 16) and the $E_{k,mean,\perp}$ equation 22:

$$\frac{\partial}{\partial t}\bar{n}k_\perp+\nabla\cdot(\bar{\Gamma}_{k_\perp})$$
$$=-\overline{\phi'([nT,gx])'}-\sigma_W\overline{\phi'(c_s n(1-\exp(\Lambda-\frac{\phi}{T_e})))'}$$
$$-\nu\overline{\phi'(n\nabla^2_\perp\omega)'}-\overline{n\mathbf{V}''_E\mathbf{V}''_E}:(\nabla\tilde{\mathbf{V}}_E)^T+\frac{\bar{\mathbf{J}}_p}{\bar{n}}\overline{n'\nabla\phi'}$$
$$+\overline{\phi'(\mathbf{V}_p\cdot\nabla n)'}+S_{k_\perp,n}, \quad (25)$$
$$\bar{\Gamma}_{k_\perp}=\bar{n}\tilde{\mathbf{V}}_E k_\perp+\overline{n\mathbf{V}''_E\mathbf{V}''^2_E}/2+\overline{\phi'\mathbf{J}'_p}, \quad (26)$$
$$S_{k_\perp,n}=\frac{1}{2}\overline{\mathbf{V}''^{*2}_E S_n}+\frac{D_n}{2}\overline{\mathbf{V}''^{*2}_E\nabla^2_\perp n}$$
$$-\frac{\sigma_N c_s}{2}\overline{n\mathbf{V}''^{*2}_E\exp(\Lambda-\frac{\phi}{T_e})} \quad (27)$$

The perpendicular transport terms (second term on LHS) and the Reynolds stress terms (fourth on RHS) in equations 22 and 25 have the same form as in hydrodynamic turbulence[18,24]. The interchange, sheath loss and viscous terms (first, second and third terms on the RHS) correspond to the pressure, the sheath loss, and the viscous stress tensor terms in a typical plasma momentum equation respectively. Comparing the $E_{k,mean,\perp}$ and the $k_\perp$ equations, it can be seen that both the Reynolds stresses and the Favre averaging term (fifth term on RHS) exchange energy between the turbulence and the mean flow. The latter originates from the non-commutative properties of Favre averaging and the divergence operator (see equation 20). Close inspection reveals that the Favre term has a structure similar to the turbulent transport and Reynolds energy transfer terms, which is in accordance with it appearing as an energy transfer term.

### C. Relation between interchange term and turbulent fluxes

The interchange term in the perpendicular energy equations will be shown to be the dominant source of kinetic energy in section IV. Hence, it is of crucial importance to model this term correctly. Interestingly, an analytical expression is found that relates this important term to the ExB energy flux. In order to more clearly demonstrate that this relation is generally applicable, this section will use equations in dimensional units (not normalised).

The interchange term in the charge balance equation or vorticity equation is the divergence of the diamagnetic current $\mathbf{J}_*=(\mathbf{B}\times\nabla p)/B^2$. Using the low $\beta$ approximation $\nabla\times\mathbf{B}\approx 0$, this is rewritten as

$$\nabla\cdot\mathbf{J}_*=\nabla\cdot(\frac{\mathbf{B}\times\nabla p}{B^2})\approx\nabla p\cdot(\nabla\frac{1}{B^2}\times\mathbf{B}). \quad (28)$$

In order to arrive at the interchange term $G_{Ek}$ in the $E_{k,\perp}$ equation, this divergence needs to be multiplied by the electric potential $\phi$:

$$G_{Ek}=-\phi\nabla\cdot\mathbf{J}_*\approx-\phi\nabla p\cdot(\nabla\frac{1}{B^2}\times\mathbf{B})$$
$$=-\nabla(p\phi)\cdot(\nabla\frac{1}{B^2}\times\mathbf{B})-p\mathbf{V}_E\cdot\nabla\ln(B^2). \quad (29)$$

An analogous derivation is carried out for the interchange source in the $k_\perp$ equation, yielding

$$G_k=-\overline{\phi'\nabla\cdot\mathbf{J}'_*}$$
$$=-\nabla(\overline{p'\phi'})\cdot(\nabla\frac{1}{B^2}\times\mathbf{B})-\overline{p'\mathbf{V}'_E}\cdot\nabla\ln(B^2). \quad (30)$$

Formula 30 is derived under the assumption of electrostatic turbulence, such that there are no fluctuations in the magnetic field $\mathbf{B}$. The relations for the interchange term reported here are similar to relations reported by Garcia et al.[26,29].

Using the low $\beta$ approximation again, the first term in expressions 29 and 30 can be rewritten as

$$-\nabla(p\phi)\cdot(\nabla\frac{1}{B^2}\times\mathbf{B})\approx-\nabla\cdot(\frac{\mathbf{b}}{B}\times\nabla p\phi)$$
$$=-\nabla\cdot(\frac{\mathbf{b}}{B}\times(\phi\nabla p+p\nabla\phi))=-\nabla\cdot(\phi\mathbf{J}_*+p\mathbf{V}_E), \quad (31)$$
$$-\nabla(\overline{p'\phi'})\cdot(\nabla\frac{1}{B^2}\times\mathbf{B})\approx-\nabla\cdot(\overline{\phi'\mathbf{J}'_*}+\overline{p'\mathbf{V}'_E}). \quad (32)$$

Hence, this first contribution to the interchange term is purely due to transport. This expression for the interchange term is identical to Scott[1]'s treatment of it. However, in a 1D geometry where the magnetic field only changes in the radial $x$-direction, it can be seen that this first contribution drops out as $\nabla_y(p\phi)$ vanishes on average.

In such a 1D case, the relation between the interchange source of kinetic energy and the ExB energy flux $5p\mathbf{V}_E/2$ becomes even more clear. Also, the second term is purely driven by the fluctuations in that case, as there is no mean radial component of the ExB flow. In this case, the interchange drive for the turbulence can hence only be positive if the ExB energy flux is in the direction of a decreasing magnetic field. This is consistent with the ballooning turbulence observed in experiments[19,29–32]. This also implies that the interchange term on the high field side acts as a sink of the turbulence (not just the absence of a source) if there is an outward ExB turbulent energy flux. In real tokamaks, this effect may be reduced or compensated by the contributions that vanish in this 1D transport model or by other sources of the turbulence. If the plasma is additionally assumed to be isothermal, relation 30 simplifies to

$$G_{k,1D,iso} = -T\bar{\Gamma}_{E,t} \cdot \nabla_x \ln(B^2). \quad (33)$$

For the TOKAM2D model presented in section II, the magnetic field gradients appear in the form of the factor $g$, so

$$G_{k,T2D} = -\overline{\phi'([p,gx])'} = g\overline{p'V'_{E,x}}, \quad (34)$$

$$G_{k,T2D,iso} = gT\bar{\Gamma}_{E,t}, \quad (35)$$

which is written in the normalised TOKAM2D units again. Relation 35 will be retrieved exactly when processing TOKAM2D simulations in section IV B. Hence, it suffices to model the turbulent ExB particle flux to model the interchange term in the $k_\perp$ equation in the considered case of an isothermal 1D transport model.

## IV. CHARACTERISATION OF THE DIFFERENT TERMS IN THE ENERGY EQUATION

In this section, we investigate the balance of turbulent kinetic energy described by equation 25. The different terms in this equation are evaluated for a set of TOKAM2D simulations to identify the dominant sources and sinks in section IV A. Then, we propose closure models for the dominant terms in Section IV B.

### A. Turbulent kinetic energy balance in 2D isothermal interchange turbulence

To assess the various terms in the $k_\perp$ equation, we performed a set of 19 TOKAM2D simulations with varying input parameters $\sigma$, $g$, $T_i$ and $\nu$. The reference simulation has parameters $\sigma = 1e-4$, $g = 6e-4$, $T_i = 1$ and $\nu = 5e-3$ as in Ref. 9. These parameters have then respectively been varied by factors $0.5-2$, $0.75-1.5$, $0.5-2$ and $0.4-3$ of the reference set. The complete set of input parameters for the simulations is provided in A.

Figures 1 and 2 show the evaluation of the different terms in the $k_\perp$ equation 25 for the default case with the standard parameter settings. Figure 1 shows the larger terms, while the smaller contributions are grouped as "other terms" in that figure and plotted separately in figure 2. In these figures, the transport terms in the left hand side of equation 25 have been moved to the right hand side (i.e. a minus sign was added to those terms). Figures 1 clearly shows that the interchange term is the dominant source of $k_\perp$ while the sheath loss term is the dominant sink. The viscous term provides a secondary sink for $k_\perp$. The other terms are much smaller than these first three. It is particularly interesting that the transport of $k_\perp$ is small (parallel and perpendicular transport of $k$ are summed in figures 1 and 2). Thus, the turbulent kinetic energy balance almost reduces to a local balance. Note also that the Reynolds stress energy transfer between the mean flow and the turbulence is very small for the cases studied here. In our simulations, this is the result of the sheath model constraining the electrostatic potential such that no significant ExB flow can develop in the diamagnetic ($y$-)direction. We expect this term to become important to describe turbulence suppression in cases with flow shear, and it will likely be required when generalizing the model presented here towards more complex setups. The Favre term (which entered the equations because Favre averaging and the divergence operator do not commute) also exchanges energy between turbulence and mean flow and appears to be larger than the Reynolds stresses. The Boussinesq term is very noisy, but its average value appears to be small. Hence, it seems to be more of a numerical artefact than a physical term, which indicates that the Boussinesq approximation made in the TOKAM2D equation set is self-consistent, as the corresponding correction term is not important. The $S_{k_\perp,n}$ term in equation 25, involving the source term and the diffusive term from the continuity equation, is observed to be very small in the considered part of the domain. The "subgrid" term represents the imbalance on the evaluation of the $k_\perp$ equation. It will be shown below that it is a numerical discretisation error, that can be reduced through grid refinement.

In figure 2, all transport terms have been plotted together. Figure 3 shows the different fluxes that contribute to this transport term separately. It is important to note that this figure plots the fluxes $\Gamma_{k_\perp}$ of the turbulent kinetic energy themselves, not their divergences that appear in equation 25. The divergence of the flux gives the transport term in the $k_\perp$ equation for all the perpendicular terms shown in figure 2. Note again that the parallel term has to be interpreted as the divergence of the parallel flux of $k_\perp$. The parallel flux itself is not available under the T2D model assumptions, and 3D turbulence studies are required to get more insight into the nature of this term. For this reason, the parallel flux of $k_\perp$ is much smaller than the other terms, whereas its contribution to transport is of the same order of magnitude as the mean flow convection term (but has the opposite sign). This figure shows that the transport of turbulent kinetic energy is dominated by mean flow convection.

It has to be remarked that the imbalance on the $k_\perp$ equation labelled subgrid model, is not negligible as it is actually the fourth largest term in this evaluation (see figure 1). Its magnitude is about 7% of the size of the interchange source in this case. A grid and time step refinement study has been conducted to verify that this error reduces with increasing refinement. It is also observed that the magnitude of $k_\perp$ increases as the grid is refined, presumably because the dissipative effect of the discretisation error is reduced. The commonly used cell sizes and time steps ($\Delta x = \Delta y = \rho$, $\Delta t = 1/\Omega$)[9,10] will be used in the remainder of this paper, despite the error that they seem to cause, because we found no significant impact on the underlying physics interpretation or saturation behavior of the turbulence at present. However, for a detailed analysis of the forward and inverse turbulence cascades[3,19,33,34], this implied subgrid model might play an important role, and requires further investigation.

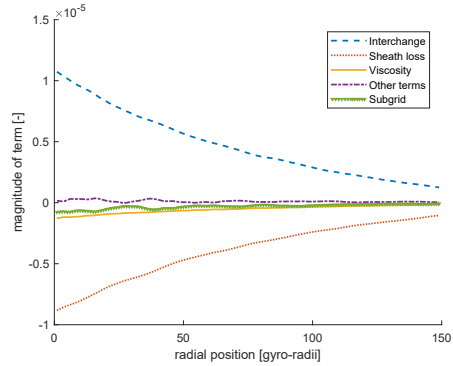

FIG. 1. Evaluation of the dominant terms in the $k_\perp$ equation 25. Other terms in the equation plotted separately in figure 2

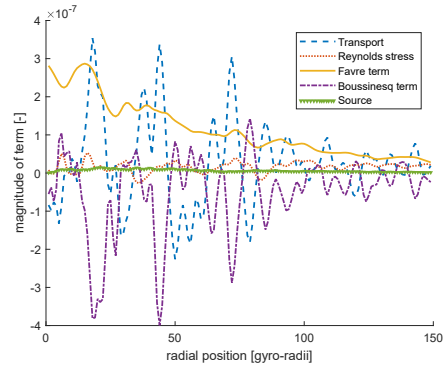

FIG. 2. Evaluation of the minor terms in the $k_\perp$ equation 25.

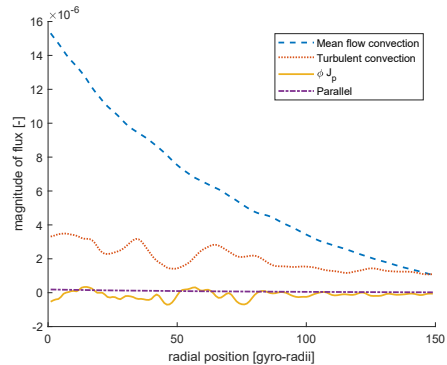

FIG. 3. Evaluation of average fluxes of $k_\perp$ in the radial and parallel direction.

### B. Development of a model for the $k_\perp$ equation

In this section, we search a model to close the $k_\perp$ equation 25. Figure 1 has shown that the interchange term, the sheath loss term and the viscous term are the dominant ones in the energy balance for $k_\perp$, which is also observed in other simulations. Hence, we focus on these terms here. First the interchange source of the turbulence will be analysed by means of a regression analysis. Then, the sheath loss and viscous sinks will be discussed, as well as the saturation mechanism of the turbulence.

#### 1. Methodology of the regression analysis

In this paper, we use non-linear least squares regression similar to that presented in Ref. 35 to extract models for the different terms in the mean-field equations that require closure. The aim is to identify the parameters $p$ in a model $f(I, p)$ that allow to reproduce the observed output quantities $O$ as closely as possible given input quantities $I$. Samples $I_i$ and $O_i$ of these quantities are available through detailed turbulence simulations. This regression analysis will also help to identify irrelevant parameters as parameters that have no significant impact on the input-output relation.

In this work, power laws are suggested as the expected model form:

$$f(I, p) = p_0 \prod_{i=1}^{N_p} I_i^{p_i}. \qquad (36)$$

In this expression, $N_p$ is the number of input quantities present in the model, which equals the number of parameters minus one (as one parameter is contained in the constant in front of the product). The exponents found in these power laws indicate whether or not a certain quantity $I_i$ in $I$ is important. This allows to trim the full set of available quantities down to those relevant for $O$.

The parameters $p$ are tuned by minimising an objective function over these parameters of the model, resulting in the

optimisation problem

$$\underset{p}{\text{minimize}} \quad \text{obj}(p) \tag{37}$$

$$\text{obj} = \sum_{i=1}^{N_d} (\frac{f(I_i, p)}{O_i} - 1)^2 \tag{38}$$

In these formulae, $N_d$ is the number of available sample points. The objective function 38 aims to minimize the relative error between model and data. The data for the regression analysis is provided by 19 TOKAM2D simulations with different combinations of the model parameters $g$, $\sigma$, $T_i$ and $\nu$ (see appendix A 2). The sample points $I_i$ and $O_i$ are the radial profiles of the relevant quantities of these simulations. To this end, the TOKAM2D data of each simulation are averaged both in time (which is allowed because only data after convergence to a statistical steady state is used) and in the $y$-direction (which is a symmetry direction).

The coefficient of determination, or $R^2$ value, will be used as a figure of merit for the models that will be tested. This value is defined as

$$R^2 = 1 - \frac{\sum_{i=1}^{N_d}(f(I_i,p) - O_i)^2}{\sum_{i=1}^{N_d}(O_i - O_{mean})^2}, \tag{39}$$

where $O_{mean} = \sum_{i=1}^{N_d} O_i/N_d$. Thus, it quantifies to what extent the variance of the data can be explained by the model.

#### 2. Regression analysis of the interchange term

The interchange term in the $k_\perp$ equation is crucial to the closure of this equation as it provides the main source of the turbulence. Subsection III C has already established the analytic relation 35 between the interchange term and the turbulent ExB energy flux. Figures 4 and 5 confirm that this relation is indeed retrieved in TOKAM2D. Figure 4 shows the radial profile of the averaged interchange term for the exact TOKAM2D data and for the analytical model 35, evaluated using TOKAM2D data, for the reference simulation with the standard parameter settings (see appendix A 1). Both lines coincide. Figure 5 shows a scatter plot of the interchange term that is obtained by evaluating relation 35 using TOKAM2D data, versus an evaluation of the exact TOKAM2D interchange term. Each data point in the figure represents a single TOKAM2D simulation (i.e. fixed parameters $g$, $T$, $\sigma$,...) that is not only averaged in time and in the diamagnetic direction, but also in the radial direction. These types of scatter plots give a clear image of the trends in the data across simulations. In addition, the coefficient of determination as introduced in equation 39 is provided (for all the data together as well as for the dataset with constant viscosity and the one with varying viscosity separately). Note that $R^2$ is calculated based on the radially averaged values plotted as data points in figure 5 and not on the full dataset of radial profiles for all simulations. The figure shows that the analytical relation 35 also manages to perfectly capture the trends in parameter space.

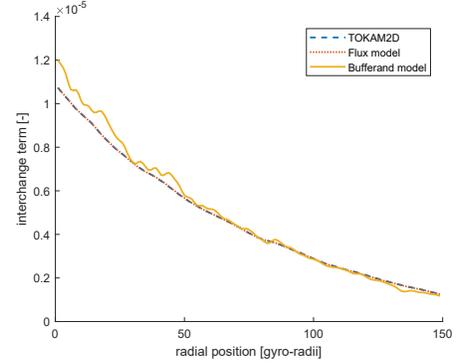

FIG. 4. Comparison of TOKAM2D interchange term with models 35 and 41.

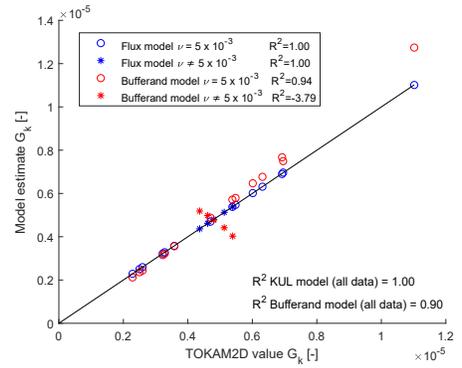

FIG. 5. Scatter plot for the interchange term models 35 and 41.

As an alternative option to model the interchange source, we analyse a model inspired by Bufferand et al.[19]. They proposed to model the interchange term using a linear growth model $G_k = \gamma \bar{n} k_\perp$, where the growth rate of the interchange instability in dimensional form is

$$\gamma = c_s \sqrt{\frac{\nabla p \cdot \nabla B}{pB} - \frac{5(1 + T_i/T_e)}{R^2}}. \tag{40}$$

Here, we adapt this model to the isothermal TOKAM2D case and remove the threshold part of the growth rate (which allowed to match the TOKAM2D data much better) yielding

$$G_k = C_{G,Buff} c_s \sqrt{-g\frac{\nabla \bar{n}}{\bar{n}}} \bar{n} k_\perp. \tag{41}$$

To determine the constant $C_{G,Buff}$, we perform a nonlinear regression on the set of TOKAM2D simulations as described in section section IV B 1. In this case, the output quantity of interest is the interchange term $O = G_k$ and $I = c_s\sqrt{-g\bar{n}\nabla\bar{n}}k_\perp$ is chosen as the explanatory variable. This procedure leads to the value $C_{G,Buff} \approx 2.21$, which is rather different from the factor $1/\sqrt{2}$ that would be obtained by filling out $g = -2\nabla B/B$ in equation 40.

The results of this alternative model are also shown in figures 4 and 5. It can be seen from figure 4 that a somewhat different (steeper) radial profile of the interchange term is obtained with model 41. The magnitude of the relative error is around 10% in this case. The default TOKAM2D case used for figure 4 lies in the middle of the parameter range investigated in the regression analysis. As a result, the model parameters are very well matched to this case specifically. Simulations further away from the center of this TOKAM2D parameter range differ more from the TOKAM2D results. This is confirmed by figure 5, which shows that Bufferand's model 41 captures the trends in parameter space rather well and has a high $R^2$ value. However, some scatter, error, remains. In particular, Bufferand's model does not seem to capture the trend for varying viscosities.

### 3. Sinks of $k_\perp$ and turbulence saturation

A regression according to section IV B 1 is conducted for the sum of the main sinks, the sheath loss and viscous terms ($O = S_k + D_k$). The explanatory quantities $I$ could be any set of quantities that are expected to be related to these terms. These may include mean-field quantities and gradients thereof (e.g. $k_\perp, n, \nabla n,...$), and TOKAM2D parameters (such as $g$, $T_i$, $\sigma$,...). The regression analysis optimises the exponents on these explanatory variables. Quantities with lower exponents have been dropped and exponents have been rounded to make the models more interpretable and physically viable. This resulted in the following model:

$$S_k + D_k = C_{sink}\sqrt{\sigma}c_s\bar{n}k_\perp \quad \text{with} \quad C_{sink} \approx -0.538. \quad (42)$$

It has to be noted that the sink is found to be proportional to $\sqrt{\sigma}c_s$, whereas it could be expected to scale as $\sigma c_s$, as that is the factor determining the strength of the sheath loss in the original vorticity equation 2. It is also worth keeping in mind that the sheath loss term $S_k$ is physically not a pure sink dissipating $k_\perp$, but is partly due to transport to the sheath.

The performance of this regression model is assessed in figures 6 and 7. Figure 6 shows that the regression model manages to capture the radial profile of the sink terms very well, with very little error remaining. The maximum relative error is smaller than 4%. The scatter plot shown in figure 7 indicates that the regression model also captures trends in parameter space rather well. Note that the $R^2$ value shown in the figure is again calculated based on the radially averaged data points shown in the figure and not on the underlying radial profiles. The largest errors here are found in simulations where the viscosity was significantly varied from its default value. The viscosity was not retained in regression model 42 because the regression analysis that was conducted showed the exponent on it to be relatively low. Another reason not to retain a scaling with the viscosity is that physically a second sink term due to viscosity would be expected, rather than a factor on the sheath loss term. This is explored in Ref. 36, where an additional equation for the turbulent enstrophy is derived to provide a measure for viscous dissipation of $k_\perp$.

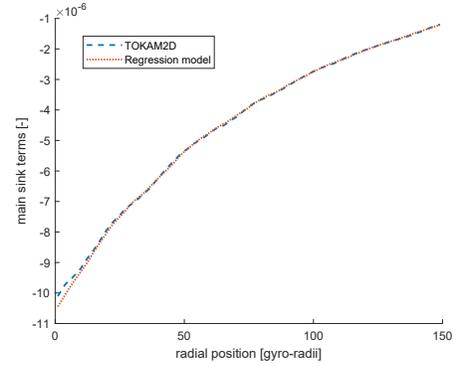

FIG. 6. Comparison of TOKAM2D sink terms with model 42.

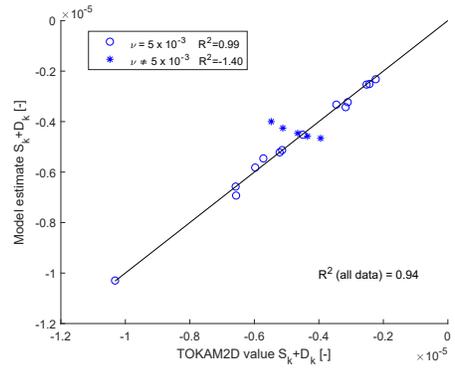

FIG. 7. Scatter plot for the sink model 42.

It is interesting to see that a purely linear sink follows from the regression analysis. This is different from the quadratic sink that was proposed in the model by Bufferand et al.. In combination with the linear source growth rate model for the interchange source model, such sink then led to a $k_\perp$ model of the form

$$\frac{\partial \bar{n}k_\perp}{\partial t} + \nabla \cdot (\Gamma_{k_\perp}) = \gamma\bar{n}k_\perp - \Delta\omega\bar{n}k_\perp^2 \quad (43)$$

In such a model, the linear drive of the turbulence causes the turbulent kinetic energy to increase initially. As $k_\perp$ increases, the nonlinear sink increases faster than the drive term and finally saturates the turbulence. Hence, the absence of any nonlinear sinks that is observed in TOKAM2D implies that the saturation mechanism is also different.

Combining the analytical relation 35 for the interchange source and regression relation 42 for the sink, we construct following model for the $k_\perp$ equation:

$$\frac{\partial \bar{n} k_\perp}{\partial t} + \nabla \cdot (\Gamma_{k_\perp}) = gT\bar{\Gamma}_{E,t} - C_{sink}\sqrt{\sigma}c_s \bar{n} k_\perp. \quad (44)$$

If steady state is assumed and the transport terms are neglected because of the quasi-local balance that is observed, the model reduces to an algebraic expression:

$$\bar{n} k_\perp = \frac{gT}{C_{sink}\sqrt{\sigma}c_s}\bar{\Gamma}_{E,t}. \quad (45)$$

The idea behind this model is the following: as soon as a turbulent ExB particle flux originates (in the direction of decreasing magnetic field, indicated by $g$), this leads to an increase in the turbulent kinetic energy (through the interchange source term), which in turn causes an increase of the particle flux (see section V) and a further build-up of the turbulence. Finally, this is saturated by a sink that is proportional to the turbulent kinetic energy. Thus, the physics of the model suggested here differ markedly from model 43, as no nonlinear sinks are required to saturate the turbulence. Note that the presented $k_\perp$ model now features a very simple model for the sink of turbulent kinetic energy, related to the parallel sheath loss term. In our model this term effectively represents $-\overline{\phi' \nabla \cdot \mathbf{J}'_\parallel}$, which can be rewritten as $-\nabla \cdot \overline{\phi' \mathbf{J}'_\parallel} + \overline{\mathbf{J}'_\parallel \cdot \nabla_\parallel \phi'}$ (see also Ref. 1). The first term can be interpreted as an additional transport term of $k_\perp$ in the parallel direction, while the second term is a real sink of $k_\perp$ related to parallel resistivity. Since we use an ad-hoc model for the parallel direction in this 2D set-up, it is impossible to distinguish between both contributions though. Hence, it is the total sheath loss term that constitutes the linear sink of $k_\perp$ that is found here for the simple sheath model used in equations 1 and 2. The term "sink" may not be completely appropriate here though as as part of this sheath loss term is thus due to transport, redistribution, of $k_\perp$ rather than dissipation. Also, we would like to stress that the parallel dynamics are likely to be more complicated than this simple linear sink. However, this work does indicate that the parallel direction suffices as a sink of the turbulence and that no nonlinear sinks (due to self saturation) are required to saturate the turbulence in the isothermal SOL case considered here. In more complex cases different saturation mechanisms might come into play though. For example, flow shear is believed to lead to turbulence quenching and may lead to zonal flow and transport barrier formation[26,28,37,38]. This phenomenon is not observed in this work as the electrostatic potential is very strongly constrained by the sheath potential such that no significant ExB flow in the diamagnetic $y$-direction and thus no flow shear can develop. Note that more complex models for the sink could easily be implemented, e.g. a second sink term to model the viscous dissipation separately, or any nonlinear terms that would be found to be important could be added in the future. It might also be envisaged to include a model for the Reynolds stresses transferring energy with zonal flow in the future. Nonetheless, the present model will be shown to explain the TOKAM2D results well already in section VI.

## V. PARTICLE TRANSPORT MODEL

In this section, models for the average radial turbulent ExB particle flux $\bar{\Gamma}_{E,t} = \overline{n' \mathbf{V}'_E}$ are developed. In Section II, we have already shown that this flux dominates the radial particle transport in the considered 1D case. Moreover, the radial mean-flow convection, which has been shown to be the dominant perpendicular transport term in the $k_\perp$ equation, is also determined by this particle flux.

The regression methodology discussed in section IV B 1 is applied to find a model for $\bar{\Gamma}_{E,t}$. The input quantities $I$ of the regression analysis are chosen as any set of quantities that are expected to be related to the particle transport (including TOKAM2D parameters, mean-field properties and gradients thereof). This yields

$$\bar{\Gamma}_{E,t} = -C_D \sqrt{k_\perp}\nabla_\perp \bar{n} \quad with \quad C_D \approx 23.9. \quad (46)$$

It has to be noted that, also here, quantities with lower exponents have been dropped and exponents have been rounded to make the models more interpretable and physically viable.

Figures 8 and 9 compare regression model 46, evaluated using TOKAM2D data, to the particle flux obtained from TOKAM2D directly. Figure 8 shows that the model captures the radial profile of the particle flux in general, however, the model profile is slightly too steep. The maximum relative error on the particle flux is 25.1%. Figure 9 shows that regression model 46 manages to capture the main trends in TOKAM2D parameter space, however, some trends seem not to be fully captured by this model. The clearest one is again the one with variations in viscosity. This time, the scaling with varying sheath loss parameter $\sigma$ does not seem to be fully captured either though. Correction factors for this have been dropped in the regression analysis as they seemed to be of secondary importance. Ref 36 has shown these errors can be significantly reduced by including the turbulent enstrophy in the diffusion coefficient. Note that the $R^2$ values shown in the figure 9 are again calculated based on the radially averaged data points shown in the figure and not on the radial profiles.

The diffusive model 46 for the average radial particle flux is very interesting in the sense that it proves to be rather robust, using a very limited number of parameters. This model indicates that the initial hypothesis that the turbulent kinetic energy $k_\perp$ plays an important role in the particle transport holds. In its dimensionless form $D = C_D \sqrt{k_\perp}$, the diffusion coefficient in this model only depends on characteristics of the turbulence. It could be argued that these are the only parameters the diffusion coefficient should depend on as the turbulence is the driver of the particle transport and no macroscopic, geometric parameters such as $g$ or $\sigma$ should be involved. Also, this square-root-scaling seems quite intuitive and is also found in hydrodynamic turbulence modelling. In RANS models for hydrodynamic turbulence, the turbulent transport of a passive scalar is often modelled using the gradient diffusion hypothesis. The diffusion coefficient therein is commonly related to the turbulent viscosity, which is assumed to scale as $l_m \sqrt{k}$ in one-equation $k$-models, where $l_m$ is a suitable length scale[18]. Recalling equation 13, in the case of 2D isothermal interchange turbulence the scaling with $\sqrt{k_\perp}$ is very natural be-cause it provides a direct measure of the local strength of the fluctuating ExB velocity.

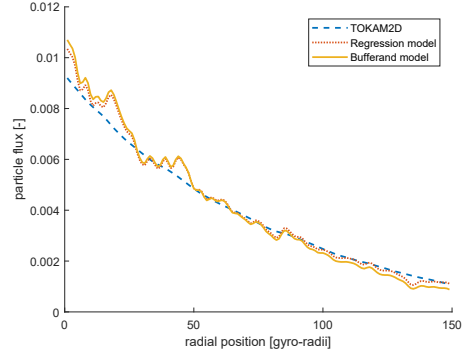

FIG. 8. Comparison of TOKAM2D particle flux with models 46 and 48.

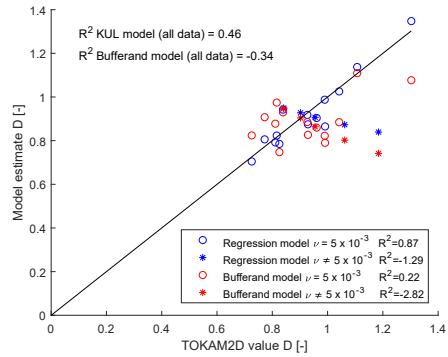

FIG. 9. Scatter plot for the diffusion coefficient of models 46 and 48.

The diffusive model proposed by Bufferand et al.[19] provides an alternative model for this particle flux:

$$\bar{\Gamma}_{E,t} \sim -\frac{ak_\perp}{c_s}\nabla \bar{n}. \quad (47)$$

In this equation, $a$ is the minor radius of the tokamak. This model is adapted to the 2D interchange turbulence model 1-3 by assuming a constant aspect ratio, such that the minor radius is proportional to the major radius, and thus inversely proportional to the magnetic field curvature $a \sim 1/g$. Thus,

$$\bar{\Gamma}_{E,t} = -C_{D,Buff}\frac{k_\perp}{c_s g}\nabla_\perp \bar{n}. \quad (48)$$

The regression methodology presented in section IV B 1 (with $O = \bar{\Gamma}_{E,t}$ and $I = k_\perp \nabla_\perp \bar{n}/(c_s g)$) yields $C_{D,Buff} \approx 0.507$. A one-to-one comparison of this constant with Bufferand et al., who used a proportionality constant of 0.67 in equation 47, is somewhat difficult because of the use of $g \sim 1/R$ in equation 48 and the minor radius $a$ in the original model 47. Trying to correct for this using $R = 2.4m$ and $a = 0.72m$ for the Tore Supra tokamak from the case ran in Ref. 19, we would find $C_{D,Buff} \approx 0.67a/R = 0.201$.

Figure 8 shows that the relative errror on the radial profile of the particle flux is more pronounced for the adapted Bufferand's model. The maximum relative error is 35.7%. Figure 9 indicates that model 48 also performs less good in parameter space.

While diffusion models for the radial particle transport are routinely used in mean-field modelling and despite the merits of the diffusive model presented here, literature seems to indicate that the nature of the particle transport in the plasma edge is not diffusive. Radially propagating structures such as avalanches and blob-filaments would rather induce intermittent convective/ballistic transport with a strong non-local character[26,39–41]. These blobs might be interpreted as a result of the gradient removal mechanism[5,6,42]. It might however be expected that a well-chosen diffusion model can give a reasonable approximation of the long time scale average particle flux caused by all the instantaneous filaments with high and low density structures moving respectively outward and inward in a seemingly random way. In addition to that, the local density gradient in a mean-field model is not just the gradient of the instantaneous background density seen by the propagating blobs, but the gradient of the time average density field including the average density of these blobs. This may partially help bridging the gap between non-local behaviour of (individual) blobs and a mean-field diffusive description. Hence, the underlying particle transport physics may not be diffusive, but its statistical average may be approximated as such to a certain extent. Note also that the transport of $k_\perp$ included in the full transport model (see equations 26 and 52) introduces a non-local effect in the mean-field model. This transport of $k_\perp$ allows turbulent kinetic energy created in one location to increase turbulent transport in another. It may nonetheless be interesting to research particle transport models that better incorporate these underlying convective properties. Inspiration for such models might be drawn from characteristic blob propagation velocity models[29,41]. Note that the main structure and interpretation of our model still hold in the case of alternative (e.g. convective) transport assumptions: the specific closure for the particle flux enters in the continuity equation, the interchange source of $k_\perp$ and the (parallel) transport of $k_\perp$, leading to a self-saturating system.

## VI. COMPLETE 1D TRANSPORT MODEL AND IMPLEMENTATION IN DIVOPT

The models developed above are combined to obtain a closed system of equations for the average density. This complete model is implemented in a 1D mean-field finite volume code that can simulate the average turbulent transport, whereas all the previous sections have only post-processed TOKAM2D data. In this section, the results of simulations

Turbulent kinetic energy in 2D isothermal interchange-dominated scrape-off layer ExB drift turbulence    12

with this model are compared to the exact TOKAM2D results, and to our interpretation of the model proposed by Bufferand *et al.*[19].

### A. Mean-field transport model for 2D isothermal interchange turbulence

Combining $k_\perp$ model 44 developed in section IV with the particle flux model 46 that depends on $k_\perp$ found in section V, the radial transport in the averaged continuity equation 8 can be closed. The resulting model equations proposed in this paper are repeated below in their dimensionless form:

$$\frac{\partial \bar{n}}{\partial t} + \nabla \cdot (\bar{n}\bar{\mathbf{V}}_E + \bar{\Gamma}_{E,t} - D_n \nabla_\perp \bar{n}) = \bar{S}_n - \sigma c_s \bar{n}, \quad (49)$$

$$\bar{\Gamma}_{E,t} = -C_D \sqrt{k_\perp} \nabla_\perp \bar{n} = -D \nabla_\perp \bar{n}, \quad (50)$$

$$\frac{\partial \bar{n} k_\perp}{\partial t} + \nabla \cdot (\bar{\Gamma}_{k_\perp}) = gT\bar{\Gamma}_{E,t} - C_{sink}\sqrt{\sigma} c_s \bar{n} k_\perp, \quad (51)$$

$$\bar{\Gamma}_{k_\perp} = k_\perp \bar{\Gamma} - C_{Dk}\bar{n} D \nabla k_\perp, \quad (52)$$

$$C_D = 23.9, \quad C_{sink} = 0.561, \quad C_{Dk} = 0.79. \quad (53)$$

Some additional assumptions have been made in the model presented above. First, a detailed study in Ref. 35 revealed that the sheath loss term in the continuity equation 49 can be approximately modelled as $\sigma c_s \bar{n}$. Moreover, an ad-hoc diffusive term has been added to the turbulent kinetic energy flux as a proxy for the (small) turbulent transport contribution in equation 52. The constant $C_{Dk}$ for the latter has been determined by means of a regression analysis. This last model is quite crude and might need to be improved in the future. It also has to be noted that the value of the $k_\perp$ sink parameter $C_{sink}$ has been determined from a regression analysis for $G_k \approx C_{sink}\sqrt{\sigma} c_s \bar{n} k_\perp$ and not as the value found in equation 42. In this way, all sinks are collected in this single sink and assumed to exactly balance the source of the turbulence locally (which is approximately observed, quasi-local balances). This allows to implicitly treat the effect of all the minor terms in the $k_\perp$ balance without having to model all of them. Note that the average particle source $\bar{S}_n$ has been included in equation 49 for generality, but that no particle source is present in the forward mean-field simulations reported in this section (since the region with a significant particle source in the TOKAM2D computational domain has been discarded in the post processing).

In the model presented here, the turbulence level and the transport are determined by the interaction between the source of the turbulence and the mean flow gradients. The source of the turbulence depends on the particle flux, which in turn depends both on $k_\perp$ and the mean flow density gradient, where the density gradient depends on the magnitude of the turbulent diffusion coefficient that is determined by $k_\perp$. Parallel sheath dynamics constitute the main sink mechanism of the turbulence.

This behaviour is compatible with the gradient removal mechanism for turbulence saturation. The idea behind this is that pressure or density gradients determine the growth rate of the turbulence (see for example equations 40 and 41). Due to the turbulence that develops, the mean flow pressure and density gradients are relaxed, leading to a reduced growth rate of the turbulence or even to its removal. This leads to intermittent behaviour, where gradients are first built up, until the instability threshold is reached and causes a sudden, large outburst due to turbulent transport[5,6,42]. Arguably, the model presented here contains the averaged result of these dynamics, which lead to a shift of the equilibrium due to the interaction between the mean-flow gradients, the resulting turbulence, and the transport caused by the turbulence. Note that the sheath losses also play an important role in the model presented here. For the simple sheath model in equations 1 and 2, the corresponding sheath loss term for $k_\perp$ constitutes a linear sink of $k_\perp$. While the parallel dynamics are likely to be more complicated than the simple models used here, these findings do indicate that the parallel direction suffices as a sink of the turbulence and that no nonlinear sinks (due to self saturation) are required to saturate the turbulence. This seems to be in accordance with Ricci and Rogers[5] and Halpern *et al.*[6], who also seem to suggest that parallel losses to the wall are the main removal mechanism for the turbulence in the gradient removal regime. More analysis of the gradient removal mechanism, and its link with the model presented here is required though. It could be especially illuminating to investigate time series and transient behaviour in TOKAM2D.

It is important to emphasise that this model is only strictly valid for the isothermal, interchange-dominated, SOL case that we considered. In this case, the electrostatic potential is to a large degree set by the sheath potential. As a result, no significant ExB flow in the diamagnetic y-direction can develop. If the electrostatic potential has more freedom to develop, strong ExB flows tend to evolve in the y-direction. In the presence of radial shear of the ExB velocity in the y-direction, these flows are maintained by extracting energy from the turbulence through the Reynolds stress terms $(\overline{n\mathbf{V}''_{E,x}\mathbf{V}''_{E,y}} : (\nabla_x \hat{\mathbf{V}}_{E,y})^T$ on the RHS of equations 22 and 25). This interaction between turbulence and ExB flows in the y-direction provides an alternative saturation mechanism for the turbulence[26,28,37,38], which is not observed in the present case, but certainly not ruled out in general. In 2D turbulence simulations of the complete edge region, including both the SOL and the edge region inside the separatrix, Garcia *et al.*[26] and Nielsen *et al.*[43] observed strong flows in the diamagnetic direction around the separatrix for example. In the future, we intend to try to integrate the interaction between turbulence and zonal flows into the $k_\perp$ transport model we developed in this manuscript. To this end, the Reynolds stress term in the exact $k_\perp$ equation 25 would need to be added to the model $k_\perp$ equation 51. This requires a closure model for both the Reynols stresses $\overline{n\mathbf{V}''_E\mathbf{V}''_E}$ themselves and a model (equation) for the mean ExB flow in the diamagnetic direction $\hat{\mathbf{V}}_{E,y}$. Manz *et al.*[38] analysed this using a mean-field versus turbulent kinetic energy approach, while Held *et al.*[28] made use of Favre averaging to further disentangle the various contributions to shear flow generation. In addition to this, the radial turbulent diffusion coefficient might also need to be adjusted for the presence of flow shear.

The new model 49-53 introduced here can be compared to the model proposed by Bufferand *et al.*, which has been adapted to the TOKAM2D case as summarised here:

$$\bar{\Gamma}_{E,t} = -C_{D,Buff}\frac{k_\perp}{mc_s g}\nabla_\perp \bar{n} = -D_{Buff}\nabla_\perp \bar{n}, \quad (54)$$

$$\frac{\partial \bar{n}k_\perp}{\partial t} + \nabla \cdot (\bar{\Gamma}_{k_\perp})$$
$$= C_{G,Buff}c_s\sqrt{-g\frac{\nabla\bar{n}}{\bar{n}}}\bar{n}k_\perp - C_{sink,Buff}\bar{n}k_\perp^2, \quad (55)$$

$$\bar{\Gamma}_{k_\perp} = k_\perp\bar{\Gamma} - C_{Dk}\bar{n}D_{Buff}\nabla(k_\perp), \quad (56)$$

$$C_{D,Buff} = 0.507, \; C_{G,Buff} = 2.21, \; C_{sink,Buff} = 5.84. \quad (57)$$

In these equations, the source term consists of the interchange term found in expression 41 and a sink term quadratic in $k_\perp$ that saturates the linear source is assumed to exist. As a crude approximation, this sink is assumed not to scale with any TOKAM2D parameters, i.e. $sink_k \sim \bar{n}k_\perp^2$ only. Note however that recent studies do include a dependence on machine parameters in this sink term in the Bufferand *et al.*'s model[21–23]. A regression analysis has been conducted to determine the corresponding constant $C_{sink,Buff}$ such that equation $\frac{C_{G,Buff}c_s}{C_{sink,Buff}}\sqrt{-g\frac{\nabla\bar{n}}{\bar{n}}}$ matches the TOKAM2D $k_\perp$ results across all simulations as good as possible. Note however that this sink term is ad-hoc as no large nonlinear sink terms have been identified in TOKAM2D in this analysis. Comparing with the value of $10^{-2}s/m^2$ reported in Ref. 19 is again not straightforward, because of the normalisation that is involved. If we assume a reference temperature of $50eV$ (temperature at the LCFS used in Ref. 19) and a magnetic field of $2T$, we would find the coefficient $C_{sink,Buff} \approx 0.25$.

### B. Assessment of mean-field model performance

The mean-field transport model presented above has been implemented in DivOpt[44], an in-house 2D (poloidal plane) finite volume code that is used for testing purposes. Zero flux boundary conditions are applied in the diamagnetic direction for both the continuity equation and the $k_\perp$ equation to arrive at a 1D code. On the radial boundaries, flux boundary conditions that exactly match the TOKAM2D data are applied for both equations. Note that we leave the development of boundary conditions for the $k_\perp$ equation for further research.

Note that the $k_\perp$ PDE model equations 51 and 55 could be simplified to algebraic $k_\perp$ equations by making a steady-state-quasi-local-balance-approximation to neglect time rate of change and the transport of $k_\perp$. This removes the need to provide boundary conditions for the $k_\perp$ equation, and has been shown to provide good results for these 1D mean-field cases. However, attention is focused on the PDE versions of the $k_\perp$ equations as it is expected that transport terms will become more important in realistic geometries such that the balance of $k_\perp$ will be non-local.

Figures 10-16 show the resulting profiles for a number of quantities of interest for two DivOpt simulations, compared to the exact TOKAM2D results. The two simulations that are compared are one with the transport model 49-53

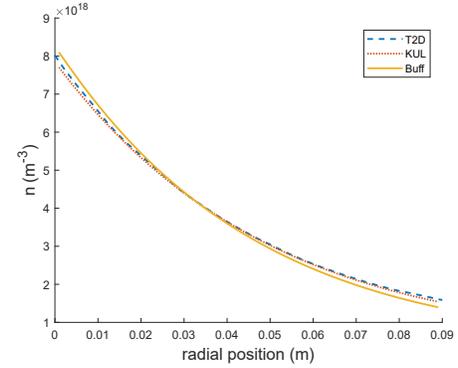

FIG. 10. Comparison of density profiles of forward DivOpt simulations with different models for $k_\perp$ and particle transport.

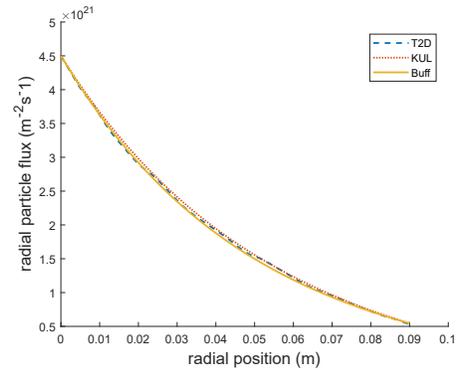

FIG. 11. Comparison of particle flux profiles of forward DivOpt simulations with different models for $k_\perp$ and particle transport.

proposed in this paper, and one with the tuned Bufferand model 54-57. The results shown here are the ones for the default TOKAM2D case. The global reference values used for making the TOKAM2D variables dimensional are $n_{ref} = 10^{19}m^{-3}$, $T_{ref} = 50eV$, $B_{ref} = 1.725T$ and $m$ is the mass of deuterium. The simulations are run on a grid with 48 cells in the radial direction.

Figures 10 and 11 show that both models are capable of predicting the density and the particle flux quite well. The "secondary quantities" being the diffusion coefficient and the turbulent kinetic energy are also approximated relatively well by the different models, as can be seen in figures 12 and 13. However, the error on these quantities is significantly higher and the difference between the various models is much more pronounced.

The KUL model proposed in this paper can be seen to capture the trend in the TOKAM2D $k_\perp$ profile very well,

Turbulent kinetic energy in 2D isothermal interchange-dominated scrape-off layer ExB drift turbulence    14

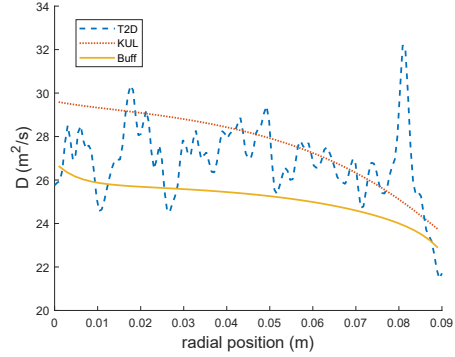

FIG. 12. Comparison of diffusion coefficient profiles of forward DivOpt simulations with different models for $k_\perp$ and particle transport.

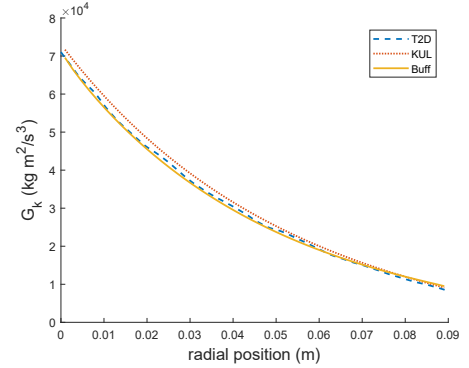

FIG. 14. Comparison of interchange source of $k_\perp$ profile of forward DivOpt simulations with different models for $k_\perp$ and particle transport.

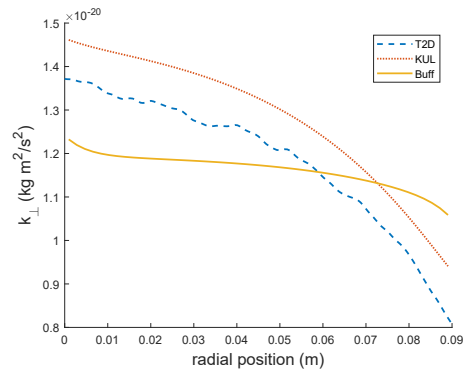

FIG. 13. Comparison of $k_\perp$ profile of forward DivOpt simulations with different models for $k_\perp$ and particle transport.

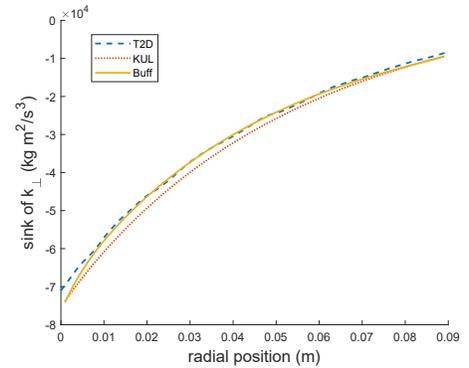

FIG. 15. Comparison of $k_\perp$ sink profile of forward DivOpt simulations with different models for $k_\perp$ and particle transport.

while the steady state Bufferand profile seems to be too flat. However, the trend in the diffusion coefficient is more similar for both models, because of the different scaling with $k_\perp$ ($D \sim \sqrt{k_\perp}$ and $D_{Buff} \sim k_\perp$ respectively). From these results we conclude that the newly developed model correctly predicts $k_\perp$, as well as its relation to the diffusion coefficient. It is hoped that including these physics will provide a good basis to extrapolate the presented model to more complex plasma flow situation. On the other hand, the model proposed by Bufferand *et al.* appears to get the trend in the diffusion coefficient right, but uses an artificial quantity to predict it, which is likely not to work anymore in more complex flow cases.

Figures 14-16 show the source, the sink and the flux of $k_\perp$. These results indicate that both models manage to realistically predict the profiles of the terms in the $k_\perp$ equation. Note that the TOKAM2D reference data used for the sink in figure 15 is that of the opposite of the interchange term, $-G_k$, as that was also the data used to fit the total sink of $k_\perp$ in equation 51.

It is worth remarking that the model parameters used in the above simulations are those tuned on the global TOKAM2D simulation set, not to this single TOKAM2D simulation in particular. However, the model parameters are very well matched to this default TOKAM2D case since it lies in the middle of the parameter range investigated in the regression analyses (see appendix A). Figure 17 on the other hand gives an idea of the kind of errors that can be expected with varying TOKAM2D parameters. It shows a scatter plot of the diffusion coefficient that is obtained by filling out the algebraic steady state local balance equivalents of $k_\perp$ models 51 and 55 respectively in the corresponding particle diffusion coefficient relations 50 and 54. This yields $D = -(C_D^2 g c_s \nabla \bar{n})/(C_{sink}\sqrt{\sigma}\bar{n})$ for the new model and $D = (C_{D,Buff} C_{G,Buff} \sqrt{\nabla \bar{n}})/(C_{sink,Buff}\sqrt{g\bar{n}})$ for

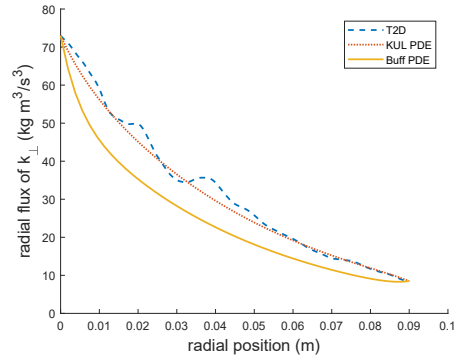

FIG. 16. Comparison of perpendicular $k_\perp$ flux profile of forward DivOpt simulations with different models for $k_\perp$ and particle transport.

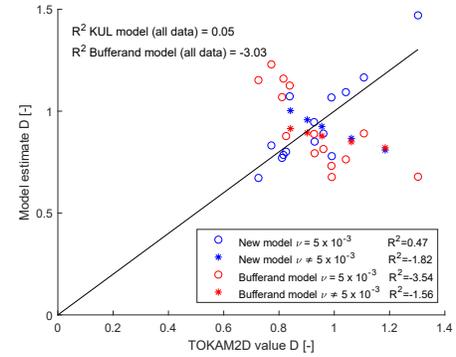

FIG. 17. Scatter plot of the predicted diffusion coefficient scaling versus the exact TOKAM2D diffusion coefficient.

Bufferand *et al.*'s model. These expressions are then evaluated using TOKAM2D data, and plotted versus the exact TOKAM2D diffusion coefficient. Each circle in figure 17 represents a single TOKAM2D simulation (i.e. fixed TOKAM2D parameters $g$, $T$, $\sigma$,...) that is not only averaged in time and in the diamagnetic direction, but also in the radial direction. Thus, this figure is constructed by post-processing TOKAM2D data, no forward DivOpt simulations have been run to make it. The $R^2$ values shown in figure 17 are again calculated based on the radially averaged data points shown in the figure. Note that if they would be calculated using the radial profiles instead, the $R^2$ values would be significantly lower because the evaluation of the radial profile introduces a large amount of noise on the model data. As the radial trend in the diffusion coefficient profile is rather weak (see blue dotted line in figure 12), the $R^2$ values would be dominated by this noise on the profile. Similar noise induced effects on the $R^2$ value were also observed in figure 9. For the data in figures 5 and 7 which have stronger radial profiles, it may be beneficial to calculate the $R^2$ on the radial profiles, but this has not been done for the sake of comparison between the different models.

Figure 17 seems to indicate that the newly proposed model manages to capture the scalings of the particle transport in TOKAM2D parameter space relatively well, although some scatter, error, does remain. The Bufferand model on the other hand appears to miss the trends in parameter space. Including a scaling with TOKAM2D parameters in the sink for $k_\perp$ in equations 55, as proposed in Refs. 21–23, might (partially) remedy this discrepancy. Upon running forward DivOpt simulations for a number of extreme values in figure 17, the newly developed model managed to still capture the trends in the profiles, although larger errors than those in figures 10-16 were observed as expected, confirming the results of 17.

## VII. CONCLUSION

This contribution has analysed the average turbulent kinetic energy in 2D isothermal electrostatic interchange-dominated ExB drift turbulence and its relation to particle transport. Models for both have been developed and calibrated, based on data from the TOKAM2D turbulence code.

A time evolution equation for the turbulent kinetic energy has been derived analytically and evaluated exactly using TOKAM2D reference data. The results indicate that the turbulent kinetic energy balance is dominated by the interchange source and a sink due to parallel current losses to the sheath. The viscous dissipation term plays a secondary role, while perpendicular transport of $k_\perp$ is observed to be small.

An analytical relation between the interchange source term of turbulent kinetic energy and the average particle flux has been derived, which directly leads to ballooning effects, especially in 1D geometries. A regression analysis has identified a model for the sheath loss sink that is linear in the turbulent kinetic energy. Moreover, a diffusive model that scales with the square root of the turbulent kinetic energy has been proposed for the average radial turbulent particle flux. Combining the latter three elements, a closed model for the radial particle flux is obtained. The transport in this model is determined by mean-flow gradients and the turbulence level, whose source is in turn controlled by the particle transport, while parallel dissipation due to the sheath suffices to saturate the turbulence. This might be interpreted as a steady state manifestation of the gradient removal mechanism for turbulence saturation[5,6,42]. It must be remarked though that zonal flows and their effect on the turbulence and the transport have not been considered in this work. Due to the isothermal conditions and the strong connection to the sheath in the parallel direction, no significant flow in the diamagnetic direction develops. If these conditions are relaxed, strong flow shear may emerge and provide an additional or alternative saturation mechanism for the turbulence (in parts of the domain) through the Reynolds stresses that are expected to become important in that case.

The developed transport model has been implemented in a 1D mean-field code and has been shown to be capable of reproducing the TOKAM2D results with high accuracy. This new model paves the way towards a significant improvement to the current best practises for modelling the radial particle transport in the SOL. The usual mean-field transport codes approximate the turbulent transport in the perpendicular directions by ad hod diffusion-type equations, which employ experimentally determined profiles for the diffusion coefficient, featuring a large amount of free parameters as the profiles are a function of the radial (and sometimes the poloidal) coordinate[13–15]. While the proposed particle transport model still features a diffusion relation, the diffusion constant now gets a clear physical background as it is shown to be determined by the energy in the ExB turbulence. The complete mean-field turbulent transport model only requires two or three (depending on whether or not the quasi-local balance approximation is used) constants to be tuned instead of a 1D or 2D field.

The results of this relatively simple model are promising and encourage its further development. Its accuracy could be further increased by the inclusion of more detailed models for the different sink and transport terms in the turbulent kinetic energy equation. The particle transport model might also still be improved; taking into account the underlying convective nature of transport might be especially interesting. The viscous dissipation term in the turbulent kinetic energy equation scales with the product of the viscosity and the enstrophy, while the scaling for the turbulent particle diffusion coefficient could also be improved by including the enstrophy[36].

The main objective of this contribution was to construct an approach and a framework for developing physics-based mean-field transport models and to apply it to a reduced turbulence model as a first step. Further research will need to extend this model to more general cases such as non-isothermal cases, cases that include the part of edge region inside the last closed flux surface and realistic tokamak geometries. Other instabilities than the interchange term as well as mean-field drift velocities might also become important then. Models for the turbulent thermal energy transport, for the new sources and/or sinks of the turbulence, as well as the additional terms present in 2D cases will need to be added.

### ACKNOWLEDGEMENT

R. Coosemans is funded by a strategic basic research grant of the Research Foundation Flanders (FWO), file number 48697.
This work has been carried out within the framework of the EUROfusion Consortium and has received funding from the Euratom research and training programme 2014-2018 and 2019-2020 under grant agreement No. 633053. The views and opinions expressed herein do not necessarily reflect those of the European Commission.
The computational resources and services used in this work were provided by the VSC (Flemish Supercomputer Center), funded by the Research Foundation Flanders (FWO) and the Flemish Government – department EWI.

### DATA AVAILABILITY STATEMENT

The data that support the findings of this study are available from the corresponding author upon reasonable request. This applies to the raw TOKAM2D turbulence code data, the mean-field data derived from it, and the mean-field data obtained from the DivOpt mean-field code.

### Appendix A: TOKAM2D simulation parameters

In all simulations the parameters were chosen such that $D_n = \nu$, $\sigma_N = \sigma_W$, $\Lambda = 2.8388$ and $Te0 = 1$. All reported simulations were run for the isothermal finite volume version version of the code with a non-periodic x-direction and using the strong Boussinesq assumption. The width of the fringe region near the inner and outer radial boundary, serving to smoothly enforce diamagnetically uniform profiles in these zones, is set to 10 gyro radii on both sides in all simulations.

#### 1. Default simulation

The default settings for the simulations are shown in table I. The last four columns are not classical TOKAM2D parameters, but are parameters used in the post-processing. $x_{start}$ and $x_{end}$ denote the first and the last cell that are considered in the post-processing. Note that these are expressed in cell number, not in gyro-radii. They serve to remove the nonphysical fringe region and the zone where the particle source is large. $t_{start}$ and $t_{end}$ denote the first and the last time steps used for the averaging, they serve to remove the non-converged first part of the simulation and to show the length of the simulation. Note that these are expressed in time steps, not in gyro-periods.

TABLE I. Default parameters used in TOKAM2D simulations and their post-processing.

| Nr. | Lx | Ly | $\Delta x$ | $\Delta y$ | $\Delta t$ | g | $Ti0$ | $\sigma$ | $\nu$ | $x_{start}$ | $x_{end}$ | $t_{start}$ | $t_{end}$ |
|---|---|---|---|---|---|---|---|---|---|---|---|---|---|
| 1 | 256 | 256 | 1 | 1 | 1 | 6e-4 | 1 | 1e-4 | 5e-3 | 51 | 199 | 2e5 | 8e5 |

#### 2. Parameter scan simulations

Table II shows the parameters of the simulations used for the regression analysis. Only the parameters that differ from the default simulation (for which the parameters are listed in table I are shown, except for the first simulation which is the default simulation. For all these simulations, the particle source $S_n$, which has a Gaussian profile in the radial $x$-direction and is uniform both in the diamagnetic $y$-direction and in time, is centered 10 gyro radii from the inner boundary and has a standard deviation of 8 gyro radii in the radial direction.

TABLE II. TOKAM2D and post-processing parameters of the simulations used in the regression analysis.

| Nr. | $g$ | $Ti0$ | $\sigma$ | $\nu$ | $t_{start}$ | $t_{end}$ |
|---|---|---|---|---|---|---|
| 1 | 6e-4 | 1 | 1e-4 | 5e-3 | 2e5 | 8e5 |
| 2 | 4.5e-4 | | | | | |
| 3 | 7.5e-4 | | | | | |
| 4 | | | 5e-5 | | | |
| 5 | | | 8e-5 | | | |
| 6 | | | 2e-4 | | | |
| 7 | | 0.5 | | | | |
| 8 | | 2 | | | | |
| 9 | | | | 2e-3 | | |
| 10 | | | | 4e-3 | | |
| 11 | | | | 6e-3 | | |
| 12 | | | | 1e-2 | | |
| 13 | | | | 1.5e-2 | | |
| 14 | 4e-4 | 0.8 | | | | |
| 15 | 4e-4 | 1.4 | | | | |
| 16 | 9e-4 | 1.8 | | | | |
| 17 | 8e-4 | 0.75 | | | | |
| 18 | | 0.9 | 0.75e-4 | | | |
| 19 | 4.5e-4 | | 1.5e-4 | | | |

### 3. Grid refinement study

A grid and time step refinement study has been conducted to verify that the error on the turbulent kinetic energy balance shown in figure 1 reduces with increasing refinement. Given the first order time integration and second order spatial discretization schemes used by the finite volumes version of the TOKAM2D code used in this contribution[9,10], the error is expected to scale like $O(\Delta x^2) + O(\Delta y^2) + O(\Delta t)$. During the refinement we systematically reduced the cell size with a factor 2 in both directions, and the time step with a factor 4 such that we would theoretically obtain second order convergence.

The exact parameters used in the TOKAM2D grid refinement simulations are shown in table III. Only the parameters that differ from the default simulation (for which the parameters are listed in table I are shown. For all these simulations, the particle source $S_n$ is centered 5 gyro radii from the inner boundary and has a standard deviation of 4 gyro radii in the radial direction.

TABLE III. TOKAM2D and post-processing parameters of the simulations used in the regression analysis.

| Nr. | Lx | Ly | $\Delta x$ | $\Delta y$ | $\Delta t$ | $x_{start}$ | $x_{end}$ | $t_{start}$ | $t_{end}$ |
|---|---|---|---|---|---|---|---|---|---|
| a | 64 | 64 | 2 | 2 | 4 | 13 | 49 | 8e5 | 1.04e7 |
| b | 128 | 128 | 1 | 1 | 1 | 26 | 99 | 2e5 | 2.6e6 |
| c | 256 | 256 | 0.5 | 0.5 | 0.25 | 51 | 199 | 8e5 | 3.2e6 |
| d | 512 | 512 | 0.25 | 0.25 | 1/16 | 101 | 399 | 1.6e6 | 8e6 |

Figure 18 shows the results for the radially averaged value of the relative error on the $k_\perp$ equation and $k_\perp$ itself as a function of the refinement step. The figure shows that the relative error and the subgrid model decrease as the grid and time step are refined. The figure also shows that the expected second order convergence has not been reached yet. Hence, the grid would have to be refined even further to complete the grid convergence study. This has not been done yet for reasons of computational cost. An alternative explanation may be that the discretization scheme we used may not be second order for one of the terms in the $k_\perp$ equation.

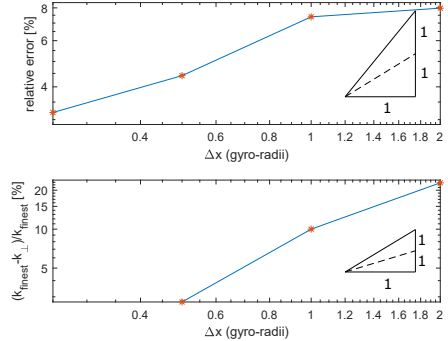

FIG. 18. Turbulent kinetic energy and relative error as a function of the number of grid cells. Time step is refined during grid refinement as well

The second plot of figure 18 shows the difference between $k_\perp$ on the finest grid and $k_\perp$ on the coarser grids. The theoretically expected second order convergence is recovered here. This figure clearly shows that $k_\perp$ increases as the grid is refined. Indeed, as the grid is refined, the dissipative effect of the discretisation error is reduced, which leads to an increase of $k_\perp$. This numerical dissipation thus acts as an additional subgrid model that is not present in the governing equations 1-3. The commonly used cell sizes and time steps ($\Delta x = \Delta y = \rho$, $\Delta t = 1/\Omega$)[9,10] will be used in the remainder of this paper, despite the error that they seem to cause, because we found no significant impact on the underlying physics interpretation or saturation behavior of the turbulence at present. However, for a detailed analysis of the forward and inverse turbulence cascades[3,19,33,34], this implied subgrid model might play an important role, and requires further investigation.

[1] B. D. Scott, Phys. Plasmas **10**, 963 (2003).
[2] J. Wesson, *Tokamaks 3rd ed.* (Oxford: Clarendon Press, 2004).
[3] W. Fundamenski, *Power Exhaust In Fusion Plasmas* (Cambridge University Press, 2010).
[4] T. T. Ribeiro and B. Scott, Plasma Phys. Contr. Fus. **47**, 1657–1679 (2005).
[5] P. Ricci and B. N. Rogers, Phys. Plasmas **20**, 010702 (2013).
[6] F. Halpern, P. Ricci, S. Jolliet, J. Loizu, and A. Mosetto, Nucl. Fusion **54**, 043003 (2014).
[7] T. Görler, X. Lapillonne, S. Brunner, T. Dannert, F. Jenko, F. Merz, and D. Told, J. Comput. Phys **18**, 7053–7071 (2011).
[8] Y. Sarazin and Ph. Ghendrih, Phys. Plasmas **5**, 4214–4228 (1998).

[9] Y. Marandet, N. Nace, M. Valentinuzzi, P. Tamain, H. Bufferand, G. Ciraolo, P. Genesio, and N. Mellet, Plasma Phys. Contr. Fusion **58**, 114001 (2016).

[10] N. Nace, *Dynamics of driven and spontaneous transport barriers in the edge plasma of tokamaks*, Ph.D. thesis, Aix-Marseille University (2018), available at http://www.theses.fr/2018AIXM0101.

[11] P. Tamain, H. Bufferand, G. Ciraolo, C. Colin, D. Galassi, Ph. Ghendrih, F. Schwander, and E. Serre, J. Comput. Phys **321**, 606–623 (2016).

[12] X. Bonnin, W. Dekeyser, R. Pitts, D. Coster, S. Voskoboynikov, and S. Wiesen, Plasma and Fusion Res. **11**, 1403102 (2016).

[13] L. Aho-Mantila, M. Wischmeier, H. Müller, S. Potzel, D. Coster, X. Bonnin, G. Conway, A. Kallenbach, L. Aho-Mantila, U. Stroth, and the ASDEX Upgrade Team, Nucl. Fusion **52**, 103006 (2012).

[14] F. Reimold, M. Wischmeier, M. Bernert, S. Potzel, D. Coster, X. Bonnin, D. Reiter, G. Meisl, A. Kallenbach, L. Aho-Mantila, U. Stroth, and the ASDEX Upgrade Team, J. Nucl. Mater. **463**, 128–134 (2015).

[15] W. Dekeyser, X. Bonnin, S. W. Lisgo, R. PITTS, D. Brunner, B. LaBombard, and J. Terry, Plasma and Fusion Res. **11**, 1403103 (2016).

[16] Y. Nishimura, D. Coster, J. Kim, and B. Scott, Contrib. Plasma Phys. **42**, 379-383 (2002).

[17] D. Zhang, Y. Chen, X. Xu, and T. Xia, Phys. Plasmas **26**, 012508 (2019).

[18] S. B. Pope, *Turbulent Flows* (Cambridge University Press, 2015).

[19] H. Bufferand, G. Ciraolo, Ph. Ghendrih, Y. Marandet, J. Bucalossi, C. Colin, N. Fedorczak, D. Galassi, J. Gunn, R. Leybros, E. Serre, and P. Tamain, Contrib. Plasma Phys. **56**, 555 – 562 (2016).

[20] K. Miki, P. H. Diamond, Ö. D. Gürcan, G. Tynan, T. Estrada, L. Schmitz, and G. Xu, Phys. Plasmas **19**, 092306 (2012).

[21] S. Baschetti, H. Bufferand, G. Ciraolo, and P. Ghendrih, J. Phys.: Conf. Ser. **1125**, 012001 (2018).

[22] S. Baschetti, H. Bufferand, G. Ciraolo, N. Fedorczak, P. Ghendrih, E. Serre, and P. Tamain, Contrib. Plasma Phys. **58**, 511–517 (2018).

[23] S. Baschetti, H. Bufferand, G. Ciraolo, N. Fedorczak, P. Ghendrih, P. Tamain, E. Serre, the EUROfusion MST1 team, and the TCV team, Nuclear Materials and Energy **19**, 200–204 (2019).

[24] V. M. Canuto, Astrophys. J. **482**, 827–851 (1997).

[25] R. Balescu, M. Vlad, F. Spineanu, and J. Misguich, Int. J. Quantum Chem. **98**, 125–130 (2004).

[26] O. E. Garcia, V. Naulin, A. H. Nielsen, and J. Rasmussen, Phys. Scr. **T122**, 89–103 (2006).

[27] T. Tran, S. Kim, H. Jhang, and J. Kim, Plasma Phys. Contr. Fusion **61**, 065002 (2019).

[28] M. Held, M. Wiesenberger, R. Kube, and A. Kendl, Nucl. Fusion **58**, 104001 (2018).

[29] O. E. Garcia, N. H. Bian, and W. Fundamenski, Phys. Plasmas **13**, 082309 (2006).

[30] J. L. Terry, S. J. Zweben, K. Hallatschek, B. LaBombard, R. Maqueda, B. Bai, C. Boswell, M. Greenwald, D. Kopon, W. Nevins, C. Pitcher, B. Rogers, D. Stotler, and X. Xu, Phys. Plasmas **10**, 1739 (2003).

[31] J. Gunn, C. Boucher, M. Dionne, I. Ďuran, V. Fuchs, T. Loarer, I. Nanobashvili, R. Pánek, J.-Y. Pascal, F. Saint-Laurent, J. Stöckel, T. Van Rompuy, R. Zagórski, J. Adámek, J. Bucalossi, R. Dejarnac, P. Devynck, P. Hertout, M. Hron, G. Lebrun, G. Moreau, F. Rimini, A. Sarkissian, and G. Van Oost, J. Nucl. Mater. **363-365**, 484–490 (2007).

[32] N. Fedorczak, J. Gunn, Ph. Ghendrih, G. Ciraolo, H. Bufferand, L. Isoardi, and P. Tamain, J. Nucl. Mater. **415**, S467–S470 (2011).

[33] A. Hasegawa and M. Wakatani, Phys. Rev. Lett. **50**, 682–686 (1983).

[34] S. Camargo, D. Biskamp, and B. Scott, Phys. Plasmas **2**, 48–62 (1995).

[35] S. Coen, *Development and validation of RANS-models for plasma edge simulations of tokamaks*, Master's thesis, KU Leuven (2018), available through Lirias: 0455596_54377855.

[36] R. Coosemans, W. Dekeyser, and M. Baelmans, Contrib. Plasma Phys. **60**, e201900156 (2020).

[37] P. Diamond, S.-I. Itoh, K. Itoh, and T. Hahm, Plasma Phys. Contr. Fusion **47**, R35 (2015).

[38] P. Manz, M. Xu, N. Fedorczak, S. Thakur, and G. Tynan, Phys. Plasmas **19**, 012309 (2012).

[39] P. A. Politzer, M. E. Austin, M. Gilmore, G. McKee, T. Rhodes, C. Yu, E. Doyle, T. Evans, and R. Moyere, Phys. Plasmas **9**, 1962–1969 (2002).

[40] Ph. Ghendrih, G. Ciraolo, Y. Larmande, Y. Sarazin, P. Tamain, P. Beyer, G. Chiavassa, X. Darmet, X. Garbet, and V. Grandgirard, J. Nucl. Mater. **390-391**, 425–427 (2009).

[41] D. A. D'Ippolito, J. R. Myra, and S. J. Zweben, Phys. Plasmas **18**, 060501 (2011).

[42] A. Mosetto, F. Halpern, S. Jolliet, J. Loizu, and P. Ricci, Phys. Plasmas **20**, 092308 (2013).

[43] A. Nielsen, J. J. Rasmussen, J. Madsen, G. Xu, V. Naulin, J. Olsen, M. Løiten, S. Hansen, N. Yan, L. Tophøj, and B. N. Wan, Plasma Phys. Contr. Fusion **59**, 025012 (2017).

[44] W. Dekeyser, D. Reiter, and M. Baelmans, Nucl. Fusion **54**, 073022 (2014).